\documentclass[12pt]{article}

\usepackage{latexsym}

\font\tenmsbm=msbm10 scaled 1200
\font\sevenmsbm=msbm9
\newfam\msbmfam
\textfont\msbmfam=\tenmsbm \scriptfont\msbmfam=\sevenmsbm

\makeatletter
\@addtoreset{equation}{section}
\makeatother

\newcommand{\eref}[1]{(\ref{#1})}

\def\be{\begin{equation}}
\def\ee{\end{equation}}
\def\ba{\begin{eqnarray}}
\def\ea{\end{eqnarray}}
\def\bet{\begin{tabular}}
\def\eet{\end{tabular}}
\def\pa{\partial}

\def\nn{\nonumber}
\def\ve{\varepsilon}

\def\a{\alpha}
\def\bt{\beta}
\def\g{\gamma}
\def\G{\Gamma}
\def\D{\Delta}

\def\dl{\delta}
\def\la{\lambda}

\def\s{\sigma}
\def\m{\mu}

\def\vp{\varphi}

\def\w{\widetilde}

\catcode`@=11

\begin{document}

\begin{titlepage}

\begin{flushright}
Preprint DFPD/2010/TH8\\
May 2010\\
\end{flushright}

\vspace{2.5truecm}

\begin{center}

{\Large \bf Quantum properties of the heterotic five--brane} \vskip0.3truecm

\vspace{2.0cm}

Kurt Lechner\footnote{kurt.lechner@pd.infn.it}

\vspace{2cm}

 {
\it Dipartimento di Fisica, Universit\`a degli Studi di Padova,

\smallskip

and

\smallskip

INFN, Sezione di Padova,

Via F. Marzolo, 8, 35131 Padova, Italia
}

\vspace{2.5cm}

\begin{abstract}

We find that the conjectured heterotic $SO(32)$ five--brane sigma model develops necessarily $k$--anomalies, and we investigate their form. We show that these anomalies can be absorbed by modifications of the superspace constraints, that satisfy automatically the modified Bianchi--identity $dH_7=X_8$ of $N=1$, $D=10$ supergravity. The $k$--anomalies induce in particular a quantum deformation of the torsion constraint $T_{\a\bt}^a=2\g^a_{\a\bt}$.

\vspace{0.5cm}

\end{abstract}

\end{center}
\end{titlepage}

\newpage

\baselineskip 6 mm
%originale
% \baselineskip 6 mm

%%%%%%%%%%%%%%%%%%%%%%%%%%%%%%%%%%%%%%%%%%%%%%%%%%%%%%%%%%%%%%%%%%%
%%%%%%%%%%%%%%%%%%%%%%%%%%%%%%%%%%%%%%%%%%%%%%%%%%%%%%%%%%%%%%%%%%%

\section{Introduction}

There are a few brane--excitations of $M$--theory that are still waiting for a low energy description in terms of a classical supersymmetric action. Examples are the systems of multiple $D$--branes \cite{My} and $M2$--branes \cite{BL}, while among the single branes the most prominent missing classical action is that of the heterotic five--brane. For the $SO(32)$ heterotic five--brane the field content has been determined in \cite{W1}, while for the $E_8\times E_8$ five--brane even the field content is still unknown. Despite the fact that these classical five--brane actions are not explicitly known, the believe in their existence is still alive. The main motivations for this credo are that they {\it exist} as excitations, in that they are the $S$--duals of the corresponding fundamental heterotic strings, and that all other five--brane excitations, the $NS5$--brane \cite{BNS}, the $D5$--brane \cite{CGNSW} and the $M5$--brane \cite{BLNPST,APPS} admit, actually, supersymmetric -- better, $k$--symmetric -- local classical actions.

The interest in generic $p$--brane $\s$--models stems from their deep relationship with the dynamics of the background field theory in which they are embedded: a brane can move consistently only in a supergravity target--space whose dynamics is properly constrained, i.e. whose fields satisfy ``effective'' equations of motion.
In the case of string $\s$--models there are several methods to derive this effective dynamics. In the supersymmetric Green--Schwarz approach it can be determined through the $\bt$--function method, relying on conformal invariance \cite{GNZ}, or, alternatively, through the $k$--anomaly method \cite{T0,T00,T000,CLT}, see below for details. These two methods are related since, in a certain sense, the square of a $k$--transformation amounts to a conformal transformation \cite{T1}. Unfortunately, the efficiency of these methods is reduced by the absence of a manifest covariant quantization procedure for the Green--Schwarz string. On the other hand, the manifestly supersymmetric pure--spinor approach \cite{B1} does entail neither $k$--symmetry nor conformal invariance, and the role of the $k$--anomaly and conformal methods is played by the requirement of nihilpotency of a certain BRST charge $Q$.

For $p$--branes with $p>1$, instead, the pure spinor approach is not available and, moreover, these objects are not conformally invariant. Under these circumstances $k$--symmetry -- and the $k$--anomaly method -- regain their fundamental roles for the derivation of the effective target--space dynamics. In this paper we enforce this method for the (conjectured) heterotic $SO(32)$ five--brane $\s$--model, to analyze its relation with $N=1$, $D=10$ supergravity in superspace, as the low energy approximation of heterotic string theory. From a phenomenological point of view this string theory appears still most promising due to the presence of non abelian gauge fields already in ten dimensions; for the $SO(32)$-- theory see e.g. \cite{P1,P2,P3}. From this point of view the heterotic five--brane gains its interest from its direct relationship with the one--loop corrected modified Bianchi identity of the seven--form,
\be\label{h7}
dH_7=\bt'X_8,
\ee
where $X_8$ is given in \eref{x8} and $1/\bt'$ is the brane tension. An open problem in string theory regards, indeed, the supersymmetrization of this Bianchi identity in superspace. For recent progress in this direction see \cite{BH} and \cite{H}. One purpose of this paper is to envisage an -- at least in principle -- systematic approach to attack this problem.

Since the low energy dynamics of the heterotic five--brane is known only in its geometric sector -- the one described by $x^m(\s)$ and $\vartheta^\m(\s)$ -- the analysis of this paper will rely essentially only on the symmetries this brane is supposed to have.
The results of the paper are summarized as follows: 1) the conjectured heterotic $SO(32)$ five--brane $\s$--model entails necessarily $k$--anomalies, whose building block we determine explicitly; 2) the cancellation of these anomalies requires a modification of the ``classical'' superspace constraints of $H_7$ and of the torsion $T^A$; 3) the Wess--Zumino consistency condition fulfilled by the $k$--anomalies ensures that the so modified $H_7$ and $T^A$ satisfy automatically the modified Bianchi identity \eref{h7} and the torsion Bianchi identity $DT^A=E^BR_B{}^A$.
A fundamental point of our analysis is that the target--space polynomial $X_8$ implies necessarily the presence of several $k$--anomalies;
4) the $k$--anomalies induce, in particular, non vanishing deformations of first order in $\bt'$ of the dimension--zero torsion,
$$
T_{\a\bt}^a=2\g^a_{\a\bt}+o(\bt'),
$$
and of the constraints of $H_7$, in agreement with \cite{H,CL}. The steps 2) and 3) represent the basic ingredients of the ``$k$--anomaly method''.

An explicit evaluation of the $k$--anomalies of the heterotic five--brane would thus lead to explicit and consistent expressions for the modified constraints. These constraints are usually ``determined''  solving superspace Bianchi identities, and we present such a ``minimal'' solution \cite{H} in section \ref{sol}, see equations \eref{ta}, \eref{hy7}, \eref{sab}, \eref{y}. The reliability of this minimal solution could thus be tested upon comparison with the constraints derived through the $k$--anomaly method. Eventually the results of this paper provide a concrete reason for why one should insist on the validity of \eref{h7} in {\it superspace}.

In a certain sense the present paper represents a generalization of the analysis of \cite{T0}, from the heterotic string to the heterotic five--brane: with this respect the main difference between strings and five--branes is that for the heterotic string the first order deformations (in that case in $\a'$) of the classical constraints of $T^A$ and $H_3$ are all {\it vanishing}.

\subsection{Gauge anomaly cancellation}

What is known about the heterotic $SO(32)$ five--brane are its gauge group ${\cal G} =SO(32)\times SU(2)$, and its $d=6$, $N=1$ supersymmetric field content \cite{W1}: the ``geometric'' sector is made out of a hypermultiplet that is singlet under ${\cal G}$, described by the fields $(x^m,\vartheta^\m)$, and the ``heterotic'' sector is made out of an $SU(2)$ Super--Yang--Mills multiplet, and of a hypermultiplet belonging to the representation $(32,2)$ of ${\cal G}$. The fermions of the hypermultiplets and of the Yang--Mills multiplet have opposite chirality in $d=6$, and the total anomaly polynomial $2\pi I_8$ has been computed in \cite{M}, see \cite{LT1} for a preliminary analysis,
\be\label{I8}
I_8=X_8+ \left(X_4+\chi_4\right)Z_4.
\ee
Here $X_8$ and $X_4$ are the standard target--space polynomials (related to the ten--dimensional Green--Schwarz anomaly polynomial through $X_{12}=X_4X_8$),
\be\label{x8}
X_8={1\over 192}\left(tr R^4+{1\over 4}\,(tr R^2)^2 -tr R^2trF^2+ 8tr F^4\right) , \quad X_4={1\over 4}\left(trR^2-trF^2\right),
\ee
and
\be
\chi_4={1\over 8}\,\ve^{ABCD}T^{AB}T^{CD}, \quad Z_4={1\over 48}\left(tr R^2-2tr T^2-24tr S^2\right).
\ee
$R,F,S$ and $T$ are  the two--form curvatures, divided by $2\pi$, associated respectively to $SO(1,9)$, $SO(32)$, $SU(2)$ and to the $SO(4)$--normal bundle. As shown in \cite{M}, the gauge anomaly \eref{I8} can be cancelled via the inflow mechanism
through the local $WZ$--term on the five--brane worldvolume,
\be
{1\over \bt'}\int B_6+ {1\over \a'} \int B_2Z_4,
\ee
where $B_6$ and $B_2$ are the dual gauge potentials of $N=1$, $D=10$ supergravity. The curvature associated to $B_2$ satisfies the Bianchi identity $dH_3=\a'X_4$.
The peculiar form of the polynomial \eref{I8}, that allows this cancellation mechanism, emerges through a number of cancellations that seem miraculous, supporting thus strongly the hypothesis that a consistent low energy effective action for the heterotic $SO(32)$ five--brane should exist.

In this paper we will be concerned only with the irreducible part of \eref{I8}, i.e. with $X_8$,  in that the interpretation of the factorized term of \eref{I8} requires presumably the knowledge of the detailed dynamics of the heterotic sector.

\section{Super five--brane $\s$--model: action and symmetries}

In this section we present the $k$--invariant super five--brane $\s$--model in a ten--dimensional supergravity background, that we consider as the building block describing the geometric sector of the heterotic five--brane. We use this model to deduce the form of the symmetries and the BRST algebra that should characterize also the heterotic five--brane. In the next section we will then use this algebra to analyze the   $\s$--model one--loop effective action of the heterotic five--brane. So, strictly speaking, the analysis of this section applies to the {\it super} five--brane.

The action of the super five--brane, rescaled by $\bt'$, is standard,
\be\label{I}
I[Z]=\int\left(\sqrt{g}\, d^6\s +B_6\right).
\ee
The supercoordinate field is $Z^M(\s)=(x^m(\s), \vartheta ^\m(\sigma))$, $m=0,\cdots,9$, $\m=1,\cdots,16,$ and $g=-det\, g_{ij}$, where the induced metric is $g_{ij}=V_i^a V_j^b\,\eta_{ab}$, and $V_i^A=\pa_i Z^M E_M{}^A(Z)$.  $A=\{a,\a\}$ stands for a ten--dimensional vector index and a sixteen--dimensional spinor index, and $i,j=0,1,\cdots,5$ are  worldvolume indices that will be raised and lowered with the metric $g_{ij}$. The second term in \eref{I} is the pullback on the worldvolume of the superspace six--form $B_6(Z)$. Target superspace zehnbein, connection, torsion and curvature are denoted respectively by $E^A=dZ^ME_M{}^A$, $\Omega_A{}^B$, $T^A=dE^A+E^B\Omega_B{}^A={1\over 2}E^BE^CT_{CB}^A$, $R_A{}^B=d\Omega_A{}^B+\Omega_A{}^C\Omega_C{}^B$.
We define the zero--order superspace constraints for $H_7=dB_6$ and $T^A$ as,
\ba
&&  \w{\ T}^{a}_{\a\bt}=2\g^a_{\a\bt}, \quad \w{\ T}^{a}_{b\a}=0, \quad \w{\ T}^\g_{\a\bt}=0,  \label{tzero}\\
&& \w H_{\a\bt a_1\cdots \,a_5}=-2(\g_{a_1\cdots \,a_5})_{\a\bt},\quad \w H_{\a a_1\cdots \, a_6}=0,\label{hzero}
\ea
where it is understood that the components of $\w H_7$ with more than two spinor indices vanish. This choice of constraints is convenient in that the dilaton $\vp$ appears in the field strength $dB_2$, but not in $dB_6$. The constraints \eref{tzero}, \eref{hzero} entail a solution of the Bianchi identities $DT^A=E^B R_B{}^A$, $dH_7=0$, that describes pure $N=1$, $D=10$ supergravity.

The equations of motion following from \eref{I} for $x^m$ and $\vartheta^\m $ respectively are,
\ba
D_iV^i_a&=&-{1\over 6! \sqrt{g}}\ \ve^{i_1\cdots \,i_6}V_{i_1}^{A_1}\cdots V_{i_6}^{A_6}\left(dB_6\right)_{aA_6\cdots A_1},\label{emx}
\\
\,[\g^j(1-\g)]^\a{}_\bt V_j^\bt&=&0. \label{emth}
\ea
The derivative $D_i$ in  \eref{emx} is covariant w.r.t. $SO(1,9)$ Lorentz--transformations and $d=6$ diffeomorphisms. In \eref{emth} we introduced the $16\times 16$ matrices,
\be
\g_i=V_i^a\g_a, \quad \g=  {1\over 6!\sqrt{g}}\ \ve^{i_1\cdots \, i_6}\,\g_{i_1\cdots \,i_6},\quad \g^2=1.
\label{gamma}
\ee
The matrices ${1\over 2}\,(1\pm\g)$ are thus projectors.

\subsection{BRST--symmetry}

The action \eref{I} is invariant under $d=6$ diffeomorphisms, with an anticommuting ghost field $c^i$, and under $k$--transformations, with a spinorial commuting ghost field $k^\a$,
\be\label{dz}
 \dl Z^M=c^i\pa_iZ^M +\D^\a E_\a{}^M,\quad\quad \D^\a\equiv {1\over2}\left(1+\g\right)^\a{}_\bt k^\bt.
\ee
While under diffeomorphisms $I[Z]$ is invariant for arbitrary background fields, under $k$--transformations it is invariant only if these fields are suitably constrained. This feature becomes manifest if one realizes that the variation of \eref{I} under \eref{dz} can be written as,
\be\label{vars}
\dl I=\int d^6\s\left(\sqrt{g}\,V^i_aV_i^B\D^\a \left(T^a_{\g B}-\w{\ T}_{\g B}^a\right)
-{1\over 6!}\ \ve^{i_1\cdots \,i_6} V_{i_1}^{A_1}\cdots V_{i_6}^{A_6} \D^\a \left(dB_6- \w H\right)_{\a A_1\cdots \,A_6}\right).
\ee
Hence, $I$ is invariant if one imposes on $T^A$ and $dB_6$ the constraints \eref{tzero}, \eref{hzero}. To derive \eref{vars} one has to use the variation,
\be\label{varV}
\dl V^A_i=c^j\pa_jV^A_i+\pa_ic^jV_j^A+ D_i\D^A  -V_i^BL_B{}^A+V_i^B\D^CT_{CB}^A,
\ee
where,
$$
\D^a\equiv 0,\quad L_a{}^b=\D^\g\Omega_{\g\, a}{}^b,\quad L_\a{}^\bt={1\over4}\,L_{ab} (\g^{ab})_\a{}^\bt, \quad L_a{}^\a=0=L_\a{}^a.
$$

It is a characteristics of $k$--transformations that the associated BRST--algebra closes only on--shell, and that it is infinitely reducible, in the sense that it requires ``ghosts for ghosts''. The transformations of the ghost fields that lead to an on--shell nihilpotent BRST--operator --  $\dl^2=0$ -- can be determined to be \footnote{Our operator $\dl$ acts from the right.},
\ba
\dl c^i&=&-c^j\pa_jc^i+\D\g^i\D,\label{dc}\\
\dl k^\a& =& -c^i\pa_ik^\a-k^\bt L_\bt{}^\a-(\D\g^i\D) V_i^\a\nn\\
&&+(V_i\g_a\D)\,[\g\g^i(\g^a-V_j^a\g^j)]^\a{}_\bt k^\bt
+{1\over2}\,(1-\g)^\a{}_\bt k_2^\bt,
\label{dk}
\ea
where $(V_i\g_a\D)\equiv V^\a_i(\g_a)_{\a\bt}\D^\bt$ etc.,
and $k_2^\bt$ indicates the second generation ghost, with ghost number two. The above transformations follow from the requirement $\dl^2Z^M=0$, upon enforcing the equation of motion \eref{emth} for $\vartheta^\m$. The presence of the term ${1\over 2}\,(1-\g) k_2$ follows essentially from the definition of the spinor $\D$ in \eref{dz}, that determines $k$ only modulo ``left--handed'' fermions. On the other hand, this term is {\it needed} in $\dl k$, because otherwise the operator $\dl$ would not square to zero on $k$. For completeness we list the transformation laws for the whole tower of ghosts $k_n$, $n=1,2,\cdots$, $k_1\equiv k$, although their explicit form will not be needed in what follows. Imposing $\dl^2 k_n=0$ for all $n$, iteratively and suppressing spinor indices, one obtains, \footnote{This recursive relation holds if one shifts the ghost $k_2$ in \eref{dk} to absorb from $\dl k$ all terms proportional to $(1-\g)$, i.e. such that $\dl k={1\over 2}\,(1+\g)\dl k+{1\over 2}\,(1-\g)k_2$.}
\be\label{dkn}
\dl k_n={1\over 2}\,\dl\g\,(k_n-\dl k_{n-1})+{1\over 2}\,(1\pm \g)k_{n+1}, \quad  n\ge 2,
\ee
where the sign is $+$ $(-)$ for $n$ even (odd). The transformation of the matrix $\g$ in \eref{gamma} is computed to be,
$$
\dl \g=
\g\,\g^i(\g_a-V^j_a\g_j)\,\dl V_i^a,
$$
where $\dl V_i^a$ is given in \eref{varV}. The above transformation laws ensure also that $\dl^2c^i=0$. Notice also the identity $V_i^a\D\g^i\D=\D\g^a\D$, implied by $\g\D=\g$.

Despite the fact that the above formulae look rather complicated, the transformation law for $\D^\a$ turns out to be rather simple, and fortunately it is all we shall need below,
\be\label{dd}
\dl \D^\a=-c^i\pa_i\D^\a-\D^\bt L_\bt{}^\a-(\D\g^i\D)V_i^\a .
\ee

\subsection{Quantization}

As anticipated above, the quantization of the super five--brane in a {\it flat} target superspace  -- as the quantization of the string in Green--Schwarz formulation -- encounters two main difficulties: 1) the BRST--algebra closes only on shell and, 2) $k$--symmetry is infinitely reducible, requiring an infinite number of ghost fields. In the case of the Green--Schwarz string these problems force eventually a non covariant quantization scheme, to truncate the infinite tower of ghosts. For a generic $p$--brane the first difficulty can, actually, be overcome employing the Batalin--Vilkovisky approach \cite{BV}, that allows to quantize systems with open gauge algebra, while the second difficulty represents still an open problem. On the contrary, for a $p$--brane in a {\it curved} target superspace, as our $\s$--model \eref{I}, one can employ the background field method combined with a normal coordinate expansion, and in this framework one can overcome both difficulties, as outlined in \cite{LT1}. More precisely, one can impose on the external classical fields the equations of motion, avoiding thus the rather cumbersome Batalin--Vilkovisky approach, and one can furthermore impose a {\it covariant} background gauge on the quantum fields, in which the infinite ghosts do not propagate. The practical outcome is that in this way one can define perturbatively  a quantum effective action $\G=I+\bt'I_1+o(\bt'^2)$, that -- in absence of anomalies -- would satisfy the on--shell relation $\dl \G=0$.

\section{$k$--anomalies}

We assume now that the quantum effective action $\G$ of the heterotic five--brane entails the BRST--symmetry constructed in the previous section. This functional defines then a local anomaly through,
\be\label{oma}
\dl\G={\cal A},
\ee
that, thanks to $\dl^2=0$, satisfies the on--shell Wess--Zumino consistency condition,
\be\label{cc}
\dl{\cal A}=0,
\ee
that will play a fundamental role in what follows.
In the following we will work at the one loop order, i.e. at first order in $\bt'$. In the rest of the paper we will show that the anomaly ${\cal A}$ is necessarily non vanishing, that \eref{cc} implies that this anomaly can be cancelled by modifying the superspace constraints \eref{tzero}, \eref{hzero} at first order in $\bt'$, and that these modifications realize a consistent solution of the superspace Bianchi identities,
\be\label{bi}
DT^A=E^BR_B{}^A, \quad\quad  d H_7=\bt'X_8.
\ee

The anomaly has to be a local functional of ghost--number one, i.e. linear in $k^\a$ and $c^i$. Since diffeomorphism anomalies can be traded for local Lorentz--anomalies, that have already been determined in \eref{I8}, it is sufficient to consider ${\cal A}$ linear in $k^\a$. Moreover, since the variations $\dl Z^M$ involve $k^\a$ only through $\D^\a$,  ${\cal A}$ is linear in $\D^\a$. In analogy with \cite{T000} we make the Ansatz,
\be\label{a12}
{\cal A}=-\int d^6\s\left(\sqrt{g}\,V^i_aV_i^B\D^\g S^a_{\g B}
+{1\over 6!}\ \ve^{i_1\cdots \,i_6} V_{i_1}^{A_1}\cdots V_{i_6}^{A_6} \D^\g W_{\g A_6\cdots A_1}\right)\equiv {\cal A}_1+{\cal A}_2,
\ee
where we introduced the targetspace superforms, of first order in $\bt'$,
\be\label{sw}
S^a={1\over 2}\,E^CE^BS^a_{BC},\quad W_7={1\over 7!}\,E^{A_7}\cdots E^{A_1}W_{A_1\cdots A_7}.
\ee
Notice that the purely vectorial components of these fields do not enter \eref{a12}.
We do not assert that \eref{a12} is the most general form of the anomaly, but we will show that on the heterotic five--brane anomalies of this kind are necessarily present.  As we will see, the forms $S^a$ and $W_7$ represent the quantum corrections to the classical constraints \eref{tzero} and \eref{hzero} for $T^a$ and $H_7$ respectively.

\subsection{Wess--Zumino consistency condition}

We impose now the Wess--Zumino consistency \eref{cc} on \eref{a12}. $\dl{\cal A}_2$ is most easily computed noting that one has ${\cal A}_2=-\int i_\D W_7$,
where $i_\D$ denotes the inner product of a superform with the vector $\D^M\equiv \D^\a E_\a{}^M$. Since ${\cal A}_2$ is invariant under $d=6$ diffeomorphisms, the variation \eref{dz} reduces to $\dl Z^M=\D^M$, that corresponds formally to a superdiffeomorphism in $D=10$, and for a superdiffeomorphism--BRST transformation on a generic $p$--superform $\Phi$ we have,
$$
\dl\left({i_\D \Phi}\right)={1\over 2}\left(i_\D i_\D d+d\, i_\D i_\D\right)\Phi.
$$
This leads to,
\be\label{var2}
\dl{\cal A}_2=-{1\over 2}\int i_\D i_\D dW_7=
-{1\over2}\int d^6\s{1\over 6!}\,\ve^{i_1\cdots \,i_6} V_{i_1}^{A_1}\cdots V_{i_6}^{A_6} \D^\a\D^\bt \left(dW_7\right)_{\a\bt A_6\cdots A_1}.
\ee
The evaluation of the variation of ${\cal A}_1$ is more cumbersome, and requires  explicit use of \eref{varV}, \eref{dd}, as well as of the equations of motion \eref{emx}, \eref{emth}. Our operator $\dl$ is nihilpotent, indeed, only on--shell \footnote{In the case of the heterotic five--brane equations \eref{emx} and \eref{emth} are necessarily modified but, as mentioned above, their use can be avoided by applying the Batalin--Vilkovisky approach.}. In particular, \eref{emx}
has to be used because $\dl V_i^\bt$, see \eref{varV}, contains $D_i\D^\bt$, and an integration by parts gives then rise to $D_iV^{ia}$. Eventually, after a rather long calculation one obtains,
\ba
\label{var1a}
\dl {\cal A}_1&=&-{1\over 2}\int d^6\s \left[3\sqrt{g}\,V^i_aV_i^C\D^\a\D^\bt
\left(D_{[C}S^a_{\a\bt)}+\w{\ T}^D_{[\a\bt}S_{DC)}^a\right)\right.\\
&&-{28\over 6!}
\,\ve^{i_1\cdots \,i_6}
V_{i_1}^{A_1}\cdots V_{i_6}^{A_6} \D^\a\D^\bt S^c_{[\a\bt} \w H_{c A_6\cdots A_1)}\label{var1b}
\Bigg].
\ea
The terms in the round brackets in \eref{var1a} are graded antisymmetrized in ${\a\bt C}$, and in \eref{var1b} there is a graded antisymmetrization over $\a\bt A_6\cdots A_1$. Eventually the total anomaly must satisfy $\dl {\cal A}_1+\dl {\cal A}_2=0$, and one could argue that the terms proportional to $\ve^{i_1\cdots \,i_6}$ and the ones proportional to $V^i_aV_i^C$ should vanish separately.
However, there can be a migration between
these two types of terms. Indeed, given an arbitrary two--superform $S^\g$ with components,
\be\label{sg}
S^\g={1\over 2}\,E^BE^C S_{CB}^\g,
\ee
one can prove the identity,
\be\label{iden}
{28\over 6!}\,\ve^{i_1\cdots \,i_6}\,
V_{i_1}^{A_1}\cdots V_{i_6}^{A_6}\D^\a\D^\bt S_{[\a\bt}^\g \w H_{\g A_6\cdots A_1)}=
3 \sqrt{g}\,V^i_aV_i^C \D^\a\D^\bt S_{[\a\bt}^D \w {\ T}_{DC)}^a.
\ee
It can be seen that this is, actually, the unique way of transforming a term of the $\ve^{i_1\cdots \,i_6}$--type into one of the $V^i_aV_i^C$--type.
Adding and subtracting \eref{iden} from \eref{var1a}, \eref{var1b}, $\dl {\cal A}= \dl {\cal A}_1+\dl {\cal A}_2$ can eventually be written as,
\ba\nn
\dl{\cal A}&=&-{1\over 2}\int d^6\s \left[3\sqrt{g}\,V^i_aV_i^C\D^\a\D^\bt
\left(D_{[C}S^a_{\a\bt)}+\w{\ T}^D_{[\a\bt}S_{DC)}^a
+ S_{[\a\bt}^D \w {\ T}_{DC)}^a
\right)\right.\\
&&-{1\over 6!}
\,\ve^{i_1\cdots \,i_6}
V_{i_1}^{A_1}\cdots V_{i_6}^{A_6} \D^\a\D^\bt\left(
28\, S^D_{[\a\bt} \w H_{D A_6\cdots A_1)} -\left(dW_7\right)_{\a\bt A_6\cdots A_1}  \right)
\Bigg].\label{varA}
\ea

\subsection{$k$--anomaly cancellation and Bianchi identities}

At first order in $\bt'$ the variation of the quantum effective action $\G=I+\bt'I_1$ amounts to,
$$
\dl \G=\dl I+{\cal A},
$$
where ${\cal A}$ is given in \eref{a12}.
The classical action $I$ of the heterotic five--brane --  that must be a completion of  \eref{I} -- is unknown. But since it must be invariant once $T^a$ and $dB_6$ satisfy the classical constraints \eref{tzero}, \eref{hzero}, we assume that $\dl I$ is still given by \eref{vars}. The requirement of anomaly cancellation $\dl \G =0$, demands then that the torsion constraints for $T^a$ have to be corrected according to,
\be\label{ta}
T^a_{\g B}=\w{\ T}^a_{\g B}+S_{\g B}^a,
\ee
and that one has to impose the constraints $\widehat H_7=\w H_7$ to the modified curvature seven--superform,
\be\label{hnew}
\widehat H_7\equiv dB_6+W_7\quad\quad \Rightarrow \quad\quad d\widehat  H_7=dW_7.
\ee
However, these new identifications for the constraints are consistent only if they satisfy the relevant Bianchi identities.
But this is guaranteed -- and this is the fundamental point -- by the consistency condition $\dl{\cal A}=0$, given \eref{varA}. Infact, defining, see \eref{sg},
\be\label{talfa}
 T^\a_{\g B}=\w {\ T}^\a_{\g B}+ S_{\g B}^\a,
\ee
and remembering that $\w {\ T}^A_{BC}$ and $\w H_7$ satisfy the Bianchi identities $D T^a=E^bR_b{}^a$  and  $dH_7=0$ at zero order in $\bt'$, up to first order in $\bt'$ \eref{varA} can be recast into,
\ba\nn
\dl{\cal A}&=&-{1\over 2}\int d^6\s \left[3\sqrt{g}\,V^i_aV_i^C\D^\a\D^\bt
\left(DT^a-E^bR_b{}^a\right)_{\a\bt C}\right.\\
&&-{1\over 6!}
\,\ve^{i_1\cdots \,i_6}
V_{i_1}^{A_1}\cdots V_{i_6}^{A_6} \D^\a\D^\bt\left(d\w H_7-
dW_7\right)_{\a\bt A_6\cdots A_1}
\Bigg].
\label{varA0}
\ea
Notice that the term involving the curvature $R_b{}^a$ does, actually, drop out from this expression, since $R_A{}^B$ is Lie--algebra valued, i.e. $R_\bt{}^a=0=R_b{}^\a$, $R_{ab}=-R_{ba}$. To examine the content of the identity $\dl{\cal A}=0$
we define the three--superform,
$$
K^a\equiv DT^a-E^bR_b{}^a={1\over 6}\,E^BE^CE^D K^a{}_{DCB}.
$$
Choosing in the first line of \eref{varA0} the index $C=\g$, the condition $\dl{\cal A}=0$ implies first of all that
$K^a$ vanishes in the sector with three spinorial indices,
\be\label{cond1}
K^a{}_{\a\bt\g}=0.
\ee
Choosing instead $C=c$, in the sector with two spinorial indices and one vector index we obtain,
\be\label{cond2}
K^{(ac)}{}_{\a\bt}=0,
\ee
because $V^i_aV_{ic}$ is symmetric in $a$ and $c$.
But, as shown in \cite{T00}, the conditions \eref{cond1}, \eref{cond2} are precisely the ones under which {\it a consistent first order solution of the whole set of Bianchi identities $DT^A=E^BR_B{}^A$ can be found}.

Similarly, defining the closed eight--superform $K_8\equiv d{\w  H}_7-dW_7$, the vanishing of the second line of \eref{varA0} implies that $K_8$ vanishes in all sectors with at least two spinorial indices,
\be\label{k2}
K_{\a\bt A_1\cdots A_6}=0.
\ee
Actually, in the sector with two spinorial and six vectorial indices the vanishing of the second line of \eref{varA0} implies,
$$
K_{\a\bt a_1\cdots a_6}=(\g^b)_{\a\bt}\,A_{b\, a_1\cdots \,a_6}+(\g_{b[a_1\cdots \, a_4})_{\a\bt}\,B^b{}_{a_5a_6]},
$$
where $A$ and $B$ are arbitrary completely antisymmetric tensors. However, the $A$--tensor can be eliminated by choosing appropriately the vectorial components $W_{a_1\cdots \,a_7}$ of $W_7$, and the $B$--tensor can be eliminated by choosing appropriately the vectorial components $S^a{}_{bc}$ of $S^a$, see \eref{sw}. This is possible because both these components do not enter into \eref{a12}. Since $K_8$ is a closed superform, \eref{k2} guarantees that $K_8$ vanishes identically, see e.g. \cite{H,LT}. This means that also the Bianchi identity  $d\widehat H_7=dW_7$ in \eref{hnew} admits a consistent solution, once one imposes on $\widehat H_7$ the constraints \eref{hzero}.

In conclusion, the consistency condition $\dl {\cal A}=0$ ensures that the modified  curvatures \eref{ta}, \eref{hnew}, \eref{talfa}, necessary for anomaly cancellation, satisfy the required superspace Bianchi identities.

\section{The $k$--anomaly of the heterotic five--brane and a coupled cohomology}

In this section we show that anomalies of the type ${\cal A}_1$ as well as those of the type ${\cal A}_2$ are necessarily present in the heterotic five--brane quantum effective action $\G=I+\bt'I_1$, and determine their form.

To this end we remember that $\G$ is also plagued by the gauge anomalies \eref{I8}, in particular by the anomaly due to the {\it target--space induced} polynomial $X_8$, that represents a gauge--anomaly ${\cal A}_G$ associated to the group  $G\equiv SO(1,9)\times SO(32)$. If we call the corresponding  nihilpotent BRST operator $\dl_G$, $\dl_G^2=0$, and use the standard transgression formalism $X_8=d\omega_7$, $\dl_G \omega_7=d\omega_6^1$, we have,
\be
 \dl_G \G ={\cal A}_G= \bt'\int \omega_6^1.\label{omg}
\ee
But since $k$--transformations preserve $G$, we have also the operatorial identity $\dl \, \dl_G=-\dl_G\, \dl$. Inserting \eref{oma} and \eref{omg} in the identity $(\dl+\dl_G)^2\,\G=0$, one obtains then the ``coupled cohomology problem'',
\be\label{coupled}
\dl{\cal A}=0,\quad \dl_G{\cal A}_G=0,\quad \dl_G{\cal A}+\dl{\cal A}_G=0.
\ee
While the first two relations are automatically satisfied, the third identity gives us new information about the $k$--anomaly ${\cal A}$. First of all, since ${\cal A}_G$ is not invariant under $k$--transformations, ${\cal A}$ is necessarily {\it non vanishing}. Moreover, ${\cal A}_G$ is known and so we can elaborate the third identity in \eref{coupled} to get a concrete information on ${\cal A}$. Since on (the pull--back of) a targetspace form $\Phi$ the $k$--transformations $\dl Z^M=\D^M $ act as the Lie--derivative $\dl \Phi=(i_\D d+d \,i_\D)\,\Phi$, we have,
\be
\dl{\cal A}_G=\bt'\int \dl \omega_6^1=\bt'\int i_\D d \omega_6^1=\bt'\int i_\D\dl_G \omega_7=\dl_G \left(\bt'\int i_\D \omega_7\right).
\ee
The third relation in \eref{coupled} becomes then $\dl_G\left({\cal A}+\bt'\int i_\D \omega_7\right)=0$, meaning that we have,
\be
{\cal A}=-\bt'\int i_\D \omega_7+{\cal A}_0,
\ee
where ${\cal A}_0$ is an ($SO(32)$ and Lorentz)--{\it invariant} $k$--anomaly. Comparing with  the general form \eref{a12} and recalling that ${\cal A}_2=-\int i_\D W_7$, we see that we must have,
\be\label{xy}
W_7=\bt' \omega_7-Y_7,
\ee
where $Y_7$ is an {\it invariant} super--form of order $\bt'$. From \eref{xy} we conclude in particular that ${\cal A}_2$ is non vanishing. Given this form of $W_7$ we may rewrite \eref{hnew} as,
\be\label{hy}
\widehat H_7+Y_7=dB_6+\bt'\omega_7, \quad\quad d(\widehat H_7+Y_7)=\bt'X_8.
\ee
We have thus derived \eref{h7}, with the identification,
\be\label{hy7}
H_7 =\widehat H_7+Y_7.
\ee
Since the constraints for $\widehat H_7$ are the classical ones given in \eref{hzero}, we see that {\it $Y_7$ represents the quantum corrections to the $H_7$--constraints}.

Given \eref{hy}, also ${\cal A}_1$ must be non--vanishing. The reason for this is that, as shown in \cite{CL2}, as long as one imposes in $D=10$, $N=1$ supergravity the dimension--zero--torsion constraint $T^a_{\a\bt}=2\g^a_{\a\bt}$, the {\it invariant} seven--superform is necessarily {\it closed}: this is in contrast to \eref{hy}, that requires instead $d(\widehat H_7+Y_7)\neq 0$. The validity of \eref{hy} implies thus that the dimension--zero--torsion in \eref{ta} gains necessarily a non--vanishing deformation $S^a_{\a\bt}$. Also ${\cal A}_1$ is, therefore, non--vanishing. Eventually, as we will see in the next section, the solution of \eref{hy} requires on top of this also a non--vanishing $Y_7$.

\subsection{Comparison with a minimal superspace solution}\label{sol}

Recently a first order ``minimal'' solution of \eref{h7}, or equivalently \eref{hy}, has been proposed in \cite{H}. The author uses a different set of superspace constraints, but since different choices of constraints are related by (more or less complicated) field redefinitions, all our conclusions above hold true. We present now a third set of constraints -- a slight ``deformation'' of the ones in \cite{H} -- for reasons that will be explained in a moment.
We replace \eref{tzero} and \eref{hzero} by,
\ba
&&  \w{\ T}^{a}_{\a\bt}=2\g^a_{\a\bt}, \quad \w{\ T}^{a}_{b\a}=0, \quad \w{\ T}^\g_{\a\bt}=2\dl_{(\a}^\g\la_{\bt)}-(\g^a)_{\a\bt}(\g_a)^{\g\dl}\la_\dl,  \label{tzero1}\\
&& \w H_{\a\bt a_1\cdots \,a_5}=-2e^{-2\varphi}(\g_{a_1\cdots \,a_5})_{\a\bt},\quad \w H_{\a a_1\cdots \, a_6}=-2e^{-2\varphi}(\g_{a_1\cdots \,a_6})_\a{}^\bt\la_\bt,  \label{hzero1}
\ea
where $\varphi$ is the dilaton, and $\la_\a=D_\a \varphi$. In this framework
\eref{I} is replaced by $I[Z]=\int(e^{-2\varphi} \sqrt{g}\, d^6\s +B_6)$, and the anomaly ${\cal A}_1$ in \eref{a12} by
${\cal A}_1=-\int(e^{-2\varphi}\sqrt{g}\,V^i_aV_i^B\D^\g S^a_{\g B})d^6\s$,
while the expression of ${\cal A}_2$ remains the same.
The main advantage of this choice consists in the fact that at zero--order in $\bt'$ it entails the ``symmetric'' parametrizations for the $SO(32)$-- and $SO(1,9)$--curvatures $F$ and $R$, see  \cite{CL2},
$$
R_{\a\bt cd}=0=F_{\a\bt}, \quad \quad R_{a\a cd}=2(\g_a)_{\a\bt}\,T^\bt_{cd},\quad F_{a\a}=2(\g_a)_{\a\bt}\,\chi^\bt,
$$
where $T^\a_{ab}$ is the gravitino field strength and $\chi^\a$ is the gluino. As a consequence the components of $X_8$ in \eref{x8} with more than four spinor indices vanish. Equation \eref{hy} is then automatically satisfied in the sectors with more than four spinor indices, if one sets the components of $Y_7$ with more than two spinor indices to zero. In the sector with four spinor indices \eref{hy} is, instead, non trivial and reads,
\be\label{44}
\left(2\g^a_{\a\bt}+S^a_{\a\bt}\right) \Big(-2e^{-2\varphi}(\g_{abcde})_{\g\dl}+Y_{\g\dl abcde}\Big)=8\,\bt'\,
(\g_{b})_{\a\vp}(\g_c)_{\bt\ve}(\g_d)_{\g\rho} (\g_e)_{\dl\s}\,C^{\vp\ve\rho\s},
\ee
where antisymmetrization over $bcde$ and symmetrization over $\a\bt\g\dl$ are understood, and we have inserted the parametrization \eref{ta} for $T^a_{\a\bt}$. The completely antisymmetric tensor $C^{\vp\ve\rho\s}$ is given by,
\be\label{c}
C^{\vp\ve\rho\s}=
tr(T^\vp T^\ve T^\rho T^\s)  +{1\over 4}\,tr (T^\vp T^\ve) tr(T^\rho T^\s) -tr(T^\vp T^\ve) tr(\chi^\rho \chi^\s)+ 8 tr(\chi^\vp \chi^\ve\chi^\rho\chi^\s),
\ee
where $T^\a$ stands for the matrix valued spinor field $T^\a_{ab}$, and antisymmetrization over $\vp\ve\rho\s$ is understood. Notice that at order zero in $\bt'$ \eref{44} is identically satisfied, while at first order in $\bt'$ it amounts exactly to the vanishing of the second line in \eref{varA} in the sector proportional to four powers of $V_i^\a$ (remember that $dW_7=\bt'X_8-dY_7$). Due to complete antisymmetry the general decomposition of \eref{c} is,
\be\label{c4}
C^{\vp\ve\rho\s}=(\g_{a_1a_2a_3})^{[\vp\ve}(\g_{b_1b_2b_3})^{\rho\s]}\left(A^{a_1a_2a_3b_1b_2|b_3}
+\eta^{a_1b_1}B^{a_2a_3|b_2b_3}\right),
\ee
where the tensors $A^{abcde|f}$ and $B^{bc|ef}$ span respectively dimension 1050 and 770 irreducible representations of $SO(1,9)$. Using
the identity,
$$
(\g_{[bc})_{(\a}{}^{(\vp}(\g_{de]})_{\bt)}{}^{\ve)}=-{1\over 12}\Big[(\g^g)_{\a\bt}(\g_{gbcde})^{\vp\ve}+(\g^g)^{\vp\ve}(\g_{gbcde})_{\a\bt}\Big]+{1\over 3}\,(\g_{bcde})_{(\a}{}^{(\vp}\dl_{\bt)}{}^{\ve)},
$$
and performing the (a bit cumbersome) gamma matrix algebra, the right hand side of \eref{44} can then be written as,
\ba
&&(\g_{b})_{\a\vp}(\g_c)_{\bt\ve}(\g_d)_{\g\rho} (\g_e)_{\dl\s}\,C^{\vp\ve\rho\s}=
{1\over 10}\,(\g_{bcdeg})_{\a\bt}(\g_{a_1a_2a_3a_4a_5})_{\g\dl}A^{a_1a_2a_3a_4a_5|g}\nn\\
&&+
(\g^g)_{\a\bt}\left[36\,(\g_{bca_1a_2a_3})_{\g\dl}A_{deg}{}^{a_1a_2|a_3}+24\,(\g^h)_{\g\dl}
A_{bcdeg|h}+16\,(\g_{bcd}{}^{a_1a_2})_{\g\dl}B_{eg|a_1a_2}\right],\nn
\ea
where (anti)symmetrizations  are understood. Consequently \eref{44} admits the solution,
\ba
\label{sab}
S^a_{\a\bt}&=&-{2\bt'\over 5} \,e^{2\varphi} \,(\g_{b_1b_2b_3b_4b_5})_{\a\bt}\,A^{b_1b_2b_3b_4b_5|a},\\
Y_{\a\bt a_1a_2a_3a_4a_5}&=&16\bt'\left(15\,(\g^{cde}{}_{[a_1a_2})_{\a\bt} A_{a_3a_4a_5]cd|e} +
6\,(\g^b)_{\a\bt}A_{a_1a_2a_3a_4a_5|b}\right. \nn\\
&&+\left.10\,(\g^{bc}{}_{[a_1a_2a_3})_{\a\bt}\,B_{a_4a_5]|bc}\right).
\label{y}
\ea
Notice that the right hand sides of these formulae are quartic in the fermions.
Formula \eref{sab} has been derived for the first time in \cite{H}, where it has also been shown that the remaining components of $S^a$, $S^\a$ and $Y_7$ are consistently determined by the Bianichi identities \eref{bi}, but their expressions are presumably much more complicated. Actually, the {\it general} solution of \eref{44} allows also for terms in $S^a_{\a\bt}$ that belong to the irreducible representations $1_a\oplus 45\oplus 54\oplus210_a$, and correspondingly for terms in $Y_{\a\bt a_1a_2a_3a_4a_5}$ that belong to $1_b\oplus 45\oplus 54 \oplus210_a\oplus210_b$. However, all these tensors can be eliminated through field redefinitions of $E^a$, $E^\a$, and $B_6$, see \cite{BB2}. From the anomaly point of view these terms can be seen to correspond to trivial $k$--anomalies, that can be absorbed subtracting local counterterms from $\G$.

If one trusts in this minimal solution, the present paper makes the testable (once a consistent formulation has been found) prediction, that the heterotic five--brane carries the $k$--anomalies \eref{a12}, where the forms $W_7$ and $S^a$ are given by \eref{xy}, \eref{sab} and \eref{y}, $S^a_{\a\bt}$ and $Y_{\a\bt a_1a_2a_3a_4a_5}$ being thus particular fourth order polynomials in $\chi^\a$ and $T_{ab}^\a$.

\section{Concluding remarks}

Whereas it is by no means clear that the low energy dynamics of the heterotic five--brane can be described by a local $\s$--model, there are at least two indications in favor of this assumption. The first is that the gauge anomalies cancel, and the second is that there exists a first order superspace solution of the associated Bianchi identity \eref{h7}. What we have shown in this paper is that a $k$--symmetric $\s$--model, together with its one--loop $k$--anomalies, are perfectly consistent with this solution. The other main reason in favor of such a model is obviously the general duality paradigm, that foresees
five--branes as $S$--duals of strings.

With respect to the heterotic string $\s$--model  \cite{T00}, from the results of our analysis the following main differences between strings and five--branes arise. For the string the $k$--anomaly is again a sum of two terms like in \eref{a12}, where $W_7$ is replaced by a three--form $W_3=\a' \omega_3-Y_3$, analogous to \eref{xy}. This leads to $H_3\equiv\widehat H_3+Y_3=dB_2+\a'\omega_3$, entailing the modified Bianchi identity $dH_3=\alpha'X_4$, where  $d\omega_3=X_4$. But in that case the three--form $Y_3$ vanished at first order in $\a'$, $Y_3=o(\a'^2)$, in that no one--loop $k$--anomaly of this kind was revealed. The $H_3$--constraints received, therefore, no corrections at first order in $\a'$. Similarly, in the case of the string also the anomaly ${\cal A}_1$ in \eref{a12} turned out to be zero at first order in $\a'$, $S^a= o(\a'^2)$, so that also the constraints for $T^a$ received no first order corrections in $\a'$. There exists, actually, a consistent all order solution of the Bianchi--identity $dH_3=\alpha'X_4$ \cite{BB}, that maintains the dimension--zero--constraint $T_{\a\bt}^a=2\g^a_{\a\bt}$ at all orders in $\alpha'$. In the case of the heterotic five--brane, instead, $Y_7$ as well as $S^a$ are already non vanishing at first order in $\bt'$.

The main open problem regarding the heterotic five--brane is the absence of a complete classical $\s$--model action, that describes also the heterotic sector. From the point of view of the present paper another open problem regards the (cancellation of the) $k$--anomaly associated necessarily -- via coupled cohomology as in \eref{coupled} -- to the factorized gauge--anomaly $(X_4+\chi_4)Z_4$. But since this anomaly is not a purely ``target--space'' anomaly, its form can be investigated probably only once also the dynamics of the heterotic sector is explicitly known.

The method presented in this paper can presumably be applied also to the $M5$--brane in $D=11$. In that case the classical action is complicated by the presence of the self--interacting chiral two--form on the worldvolume, but this time, in addition to the anomaly polynomial $X_8^{(M5)}$ \cite{W2}, also the complete $k$--invariant classical action is known \cite{BLNPST,APPS}. This allows for a ``true'' analysis, and no longer for only a ``conjectured'' one, that could in particular shed new light on non--minimal $N=1$, $D=11$ supergravity \cite{YH}.

\vskip0.5truecm

\paragraph{Acknowledgements.}
This work is supported in part by the INFN Special Initiative TV12.

\end{document}